\documentclass[manuscript, letterpaper]{aastex6}

\usepackage{graphicx}
\usepackage[suffix=]{epstopdf}
\usepackage{natbib}
\usepackage{amsmath}
\usepackage{url}
\usepackage{xspace}

\makeatletter % you know you are living your life wrong when you need to do this
\long\def\frontmatter@title@above{
\vspace*{-\headsep}\vspace*{\headheight}
\noindent\footnotesize
{\noindent\footnotesize\textsc{\@journalinfo}}\par
{\noindent\scriptsize Preprint typeset using \LaTeX\ style AASTeX6\\
With modifications by David W. Hogg and Daniel Foreman-Mackey
}\par\vspace*{-\baselineskip}\vspace*{0.625in}
}%
\makeatother
\makeatletter
\let\origsection\section
\renewcommand\section{\@ifstar{\starsection}{\nostarsection}}
\newcommand\nostarsection[1]{\sectionprelude\origsection{#1}}
\newcommand\starsection[1]{\sectionprelude\origsection*{#1}}
\newcommand\sectionprelude{\vspace{1em}}
\let\origsubsection\subsection
\renewcommand\subsection{\@ifstar{\starsubsection}{\nostarsubsection}}
\newcommand\nostarsubsection[1]{\subsectionprelude\origsubsection{#1}}
\newcommand\starsubsection[1]{\subsectionprelude\origsubsection*{#1}}
\newcommand\subsectionprelude{\vspace{1em}}
\makeatother
\sloppypar
\journalinfo{ApJ Accepted: Nov 21, 2016}

\newcommand{\Kepler}{\textsl{Kepler}\xspace}

\begin{document}

%%%%%%%%%%%%%%%%%%%%%%
\title{Rotating Stars from \Kepler Observed with Gaia DR1}

\shorttitle{Rotating Stars}
\shortauthors{Davenport}

\author{
	James R. A. Davenport\altaffilmark{1,2}
	}

\altaffiltext{1}{Department of Physics \& Astronomy, Western Washington University, 516 High St., Bellingham, WA 98225, USA}
\altaffiltext{2}{NSF Astronomy and Astrophysics Postdoctoral Fellow}

%%%%%%%%%%%%%%%%%%%%%%%%%%%%%%
\begin{abstract}
Astrometric data from the recent Gaia Data Release 1 has been matched against the sample of stars from \Kepler with known rotation periods. A total of 1,299 bright rotating stars were recovered from the subset of Gaia sources with good astrometric solutions, most with temperatures hotter than 5000 K. From these sources, 894 were selected as lying near the main sequence using their absolute $G$-band magnitudes. These main sequence stars show a bimodality in their rotation period distribution, centered roughly around a 600 Myr rotation-isochrone. This feature matches the bimodal period distribution found in cooler stars with \Kepler, but was previously undetected for solar-type stars due to sample contamination by subgiants.
A tenuous connection between the rotation period and total proper motion is found, suggesting the period bimodality is due to the age distribution of stars within $\sim$300pc of the Sun, rather than a phase of rapid angular momentum loss.
This work emphasizes the unique power for understanding stellar populations by combining temporal monitoring from \Kepler with astrometric data from Gaia.
\end{abstract}

%%%%%%%%%%%%%%%%%%%%%%%%%%%%%%
\section{Introduction}

The \Kepler mission \citep{borucki2010} has enabled the first studies of rotation periods for large ensembles of field stars. The fundamental stellar property of rotation has been measured for over 30k stars using the high cadence \Kepler light curves, tracing the periodic or quasi-periodic modulations in brightness as cool starspots rotate in and out of view \citep{reinhold2013,mcquillan2014}. The seminal work by \citet{skumanich1972} connected stellar rotation and age via angular momentum loss, leading to an age estimating technique known as gyrochronology. At present, ages determined by gyrochronology are accurate to $\sim$10\% in the {\it best} cases (young solar-type stars). Determining robust ages for field stars may soon be possible by calibrating gyro-isochrones to stellar clusters and asteroseismic samples, and improved models of angular momentum loss \citep[e.g.][]{angus2015,van-saders2016}.

%tgas = J/ApJS/211/24/table1 \& I/337/tgas
%gaia = J/ApJS/211/24/table1 \& I/337/gaia
\begin{figure*}[]
\centering
\includegraphics[width=6in]{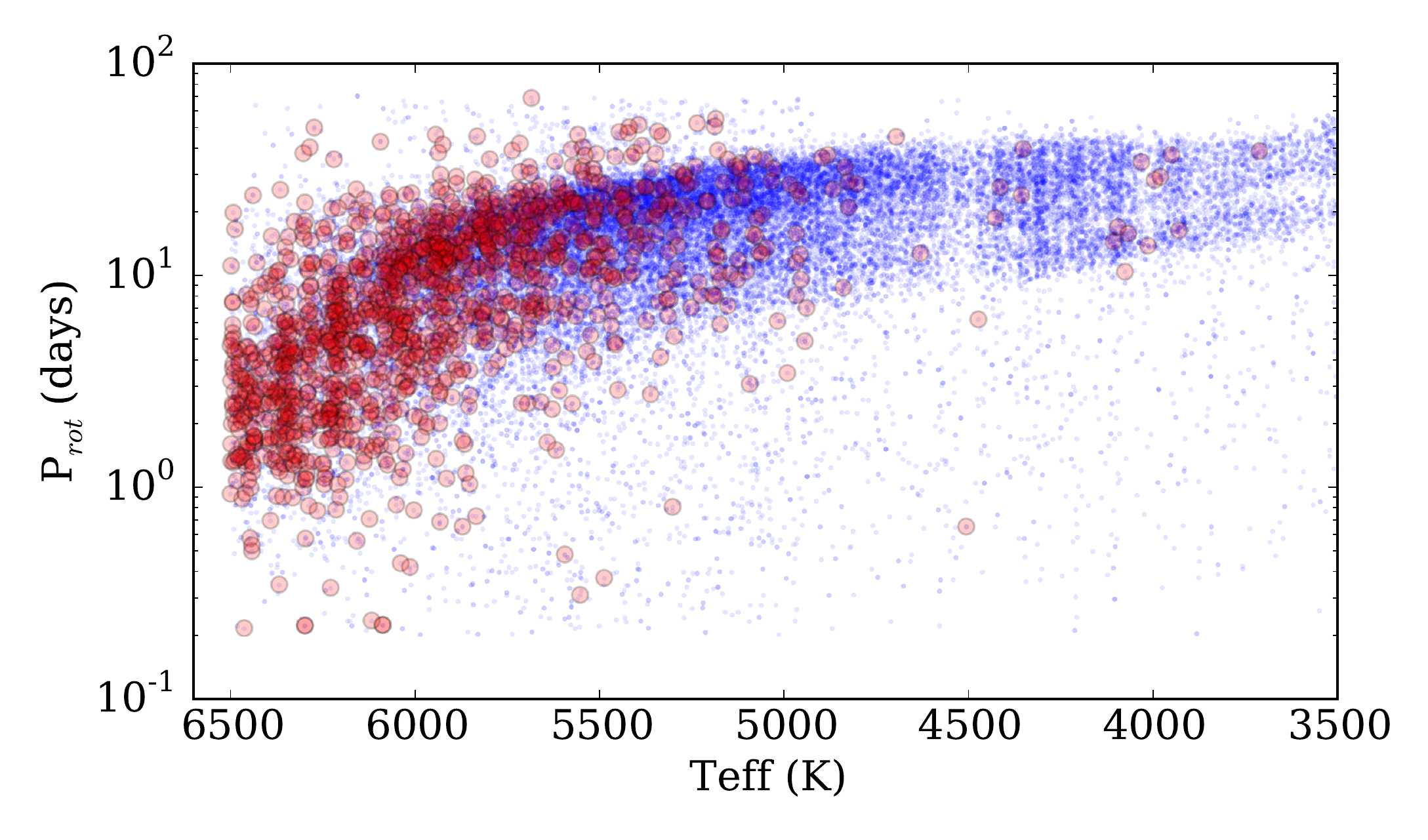}
\caption{
Rotation period distribution for 33,855 \Kepler stars from \citet{mcquillan2014} with detections in Gaia DR1 (blue dots). The period bimodality can be seen most clearly for stars with $T_{eff} < 4000$ K as a dearth of sources with periods of $\sim$25 days, but extends to at least $T_{eff}\sim5500$ K according to \citet{mcquillan2014}. The subsample of 1,299 nearby objects found in TGAS are highlighted (red circles), and are mostly hotter stars due to the faint limit of the TGAS sample. 
}
\label{fig:all}
\end{figure*}

\citet{mcquillan2013} discovered a bimodal period distribution for M dwarfs in the \Kepler field, which was subsequently confirmed to exist for K dwarfs as well \citep{mcquillan2014}. However, this bimodality had never been observed in any other study of stellar rotation periods, including stellar clusters at a variety of ages, nor was it detected in the \Kepler stars at hotter temperatures ($T_{eff} > 5000$). While binary stars and multiple-period systems may be contaminating the rotation period sample for \Kepler field stars, \citet{mcquillan2014} found the presence of such interlopers could not adequately explain the bimodal period distribution. Currently favored explanations for this feature are 1) a non-continuous age distribution for nearby stars, as was suggested with very nearby Hipparcos stars by \citet{hernandez2000}, or 2) a previously unknown phase of rapid angular momentum loss for low-mass stars, similar to the ``Vaughan-Preston'' gap seen in chromospheric activity indicators \citep{vaughan1980}. As independent age indicators for these field stars are often non-existent, and both scenarios deal with physical mechanisms that are not currently understood with precision, a definitive explanation has not been found.

Astrometric data from the Gaia mission \citep{gaia} can help shed light on this stellar population mystery. By measuring distances via stellar parallax for these rotating stars, the \Kepler--Gaia sample can separate single main sequence dwarfs from binary stars or evolved stars such as subgiants, and will help calibrate fundamental properties of \Kepler stars, such as log(g) \citep{creevey2013}. Galactic kinematics from Gaia will also provide an additional age-proxy, and allow for searches of substructure in field star ages such as from moving groups. The Gaia data will also enable a measurement of the star formation history of the disk from both white dwarf cooling sequences \citep{carrasco2014,gaensicke2015} and color-magnitude diagram models \citep{bertelli1999}.

In this paper I demonstrate the utility of combining temporal properties derived from \Kepler light curves with the preliminary astrometric solutions from  Gaia Data Release 1 \citep[hereafter DR1][]{gaia_dr1}. This combined sample allows improved selection of main sequence stars, and reveals previously undetected structure in the rotation period distribution for solar-type stars.

%%%%%%%%%%%%%%%%%%%%%%
\section{The \Kepler--Gaia Data}
Rotation periods in this study come from \citet{mcquillan2014}, who performed an Auto-Correlation Function analysis of \Kepler stars cooler than 6500 K that had at least $\sim$2 years of observation. The periods recovered from this approach generally agree very well with those found via Lomb-Scargle Periodograms \citep[e.g.][]{reinhold2013,aigrain2015}. Sources with multiple distinct periods, such as from binary systems with two spotted stars \citep[e.g.][]{lurie2015} are detected by \citet{mcquillan2014}, but are not included in the following analysis.

The Gaia Data Release 1 (DR1) provides astrometric positions for over $10^9$ sources from the first year of observation with Gaia. The Tycho-Gaia Astrometric Solution (TGAS) measures improved proper motions and parallaxes for 2 million nearby, bright sources by extending the astrometric solutions from Tycho and Hipparcos. While the TGAS data are not a complete astrometric survey, and have possible systematics in the reported parallaxes \citep{gaia_dr1,stassun2016}, they represent a significant improvement in the astrometry and kinematics available for stars in the \Kepler field.

Using the CDS X-Match service, I cross-matched the available catalogs from these two surveys. A default cross-match radius of 5 arcseconds was used. A total of 33,855 stars were found in the cross match between these catalogs, 99.5\% of the sample from \citet{mcquillan2014}. The small number of stars not recovered from \citet{mcquillan2014} may be due to source confusion within the matching radius. A subset of 1,299 objects were recovered in the TGAS sample. Due to the brightness limits of the TGAS sample very few K and M dwarfs were recovered in the TGAS sample. Future releases of Gaia data will provide full astrometric solutions for nearly all \Kepler stars. The rotation periods versus stellar effective temperatures for the \Kepler--Gaia matched stars are shown in Figure \ref{fig:all}.

%%%%%
\begin{figure}[]
\centering
\includegraphics[width=5in]{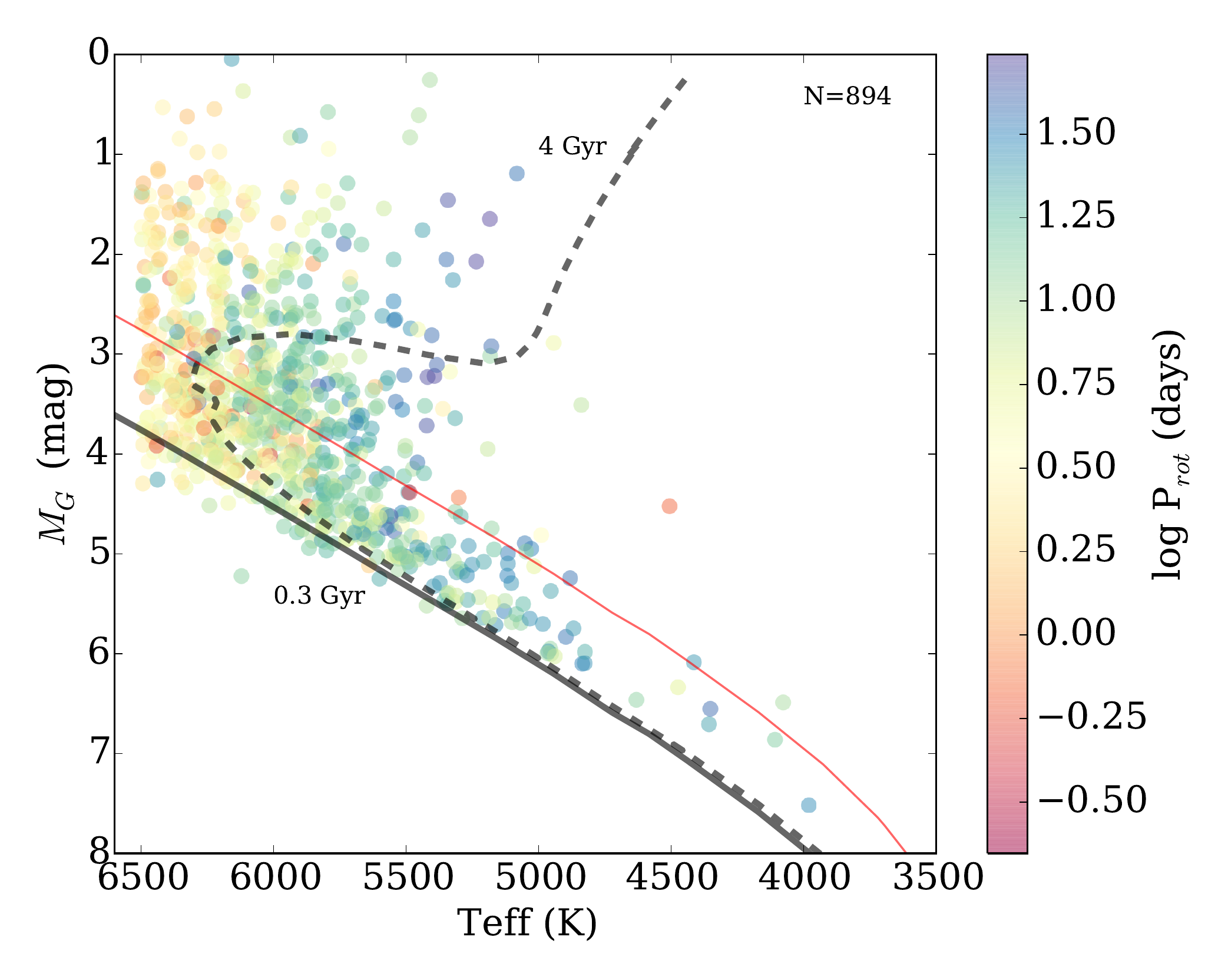}
\caption{Hertzsprung--Russell (HR) diagram using temperatures from \citet{mcquillan2014} and Gaia DR1 $G$-band absolute magnitudes for the 894 stars that pass photometric and parallax quality cuts described in the text. Points are colored by their measured \Kepler rotation periods. Two isochrones from \citet{bressan2012} are shown, with ages of 300 Myr and 4 Gyr (solid and dashed lines). Main sequence stars are selected as those between the 300 Myr isochrone and the isochrone shifted up by 1 magnitude (red line).
}
\label{fig:HR}
\end{figure}

%%%%%%%%%%%%%%%%%%%%%%
\section{Selecting Main Sequence Stars}

Though \citet{mcquillan2014} attempted to only measure periods for dwarf stars, the sample of \Kepler--Gaia matched stars contains both main sequence dwarfs and evolved stars (giants and subgiants). Previous studies have shown  significant contamination by giants or subgiants can affect the implied variability properties of dwarf stars \citep{ciardi2011,mann2012}. Therefore to properly understand the nature of the period distribution and its implications for age-dating field stars, a robust sample of main sequence stars must be selected.

Faint stars were removed by requiring sources have $G$-band flux errors $<1$\%. To ensure accurate distances, and therefore luminosities, parallaxes were required to have errors $<0.4$ mas. These cuts left a total of 894 stars from the \Kepler--TGAS matched sample (68\%). The Hertzsprung--Russell (HR) diagram for these stars is shown in Figure \ref{fig:HR}, with each point colored by its \Kepler-measured rotation period. Example isochrones from the \citet{bressan2012} grid are shown for two ages. A systematic offset of $\sim$0.5 magnitudes is found between the measured absolute $G$-band and the isochrone's main sequence. This offset is likely due to calibration differences between the nominal and actual $G$-band (A. Brown, private communication).

The HR-period diagram in Figure \ref{fig:HR} shows stars with a range of evolutionary states, and could help test post-main sequence angular momentum evolution models \citep[e.g.][]{donascimento2012}. Outliers in this diagram are either due to erroneous cross-matching in the \Kepler and Gaia catalogs, or represent interesting systems such as rare binary star configurations or stars that have undergone mergers or ingested giant planets \citep{massarotti2008,tayar2015}. For example, examining the 2 Micron All Sky Survey \citep[2MASS][]{2mass} image on SIMBAD for 
the rapidly rotating star at $T_{eff}=4500$ K, $M_G=4.5$ mag, $P_{rot}$=0.652 d, in Figure \ref{fig:HR} (KIC 07957709), this source appears highly contaminated with three point sources clustered within $\sim$10 arcseconds, likely leading to an erroneous position in the HR diagram, and possibly incorrect variability measurements with \Kepler.
Investigating all such outliers in Figure \ref{fig:HR} is beyond the scope of this work, but these targets are worth further study as they may reveal new physics.

Main sequence stars were selected using a simple cut around 300 Myr isochrone. Given the systematic offset of $\sim$0.5 mag between the Myr isochrone and the observed $M_G$ values, a fairly wide band of stars ($0\le \Delta M_G \le 1$) was selected as being ``close to the main sequence''. The final sample included 440 stars. This simplistic cut is not a robust dwarf--giant, nor single--binary star separator, but serves to select a sample of mostly main sequence stars for the illustrative purpose of this work. More precise selection will require an improved isochrone track, as well as updated parallaxes from the full Gaia DR2.

%%%%%
\begin{figure*}[]
\centering
\includegraphics[width=6in]{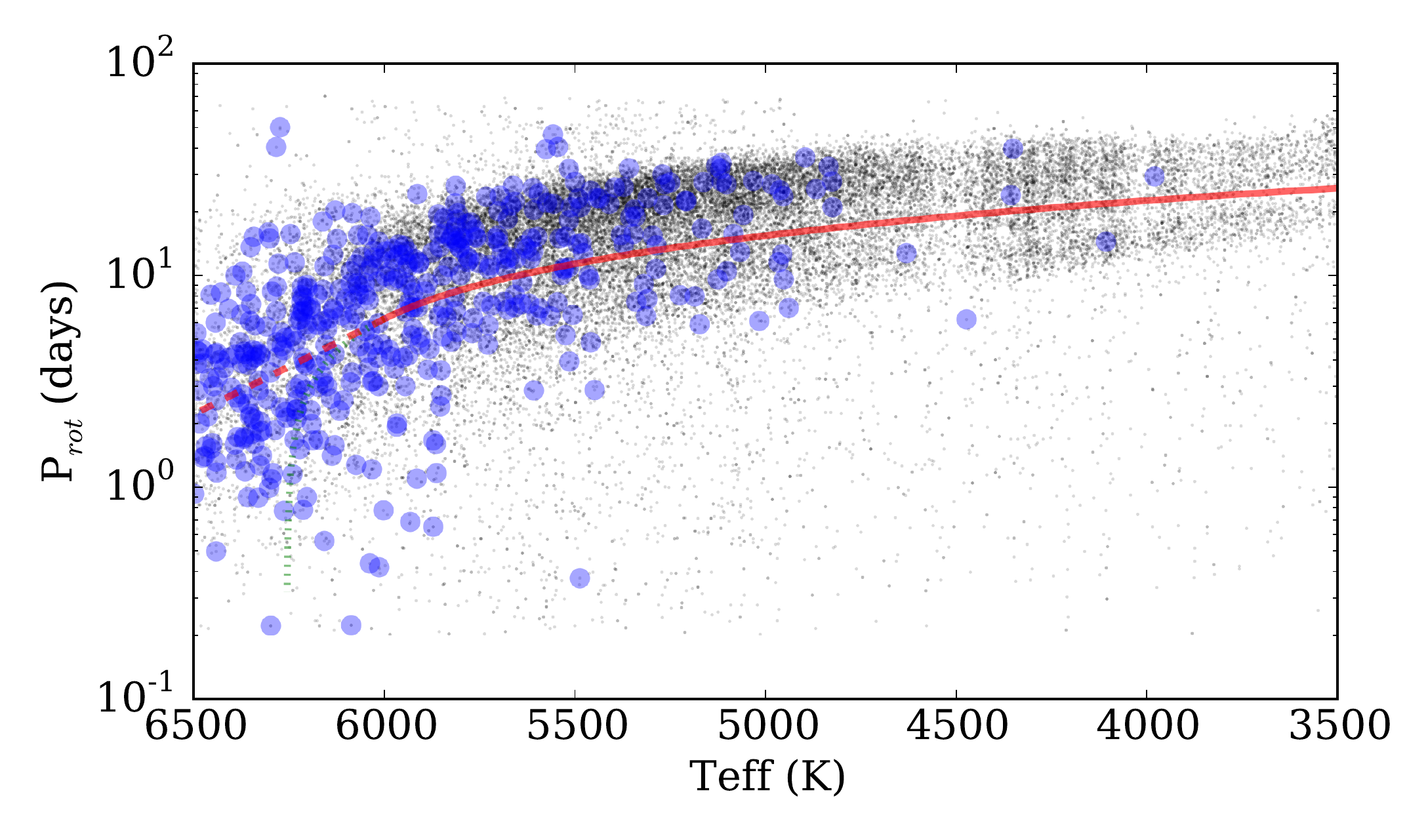}
\caption{ Rotation period versus temperature for TGAS-matched stars near the isochrone main sequence (blue circles). 
The full \Kepler--Gaia matched sample is shown for reference (small black dots). A bimodality in rotation periods initially discovered for M dwarfs by \citet{mcquillan2013}, extends the full range of temperatures in the \Kepler--Gaia main sequence sample shown here.
A \citet{meibom2011} 600 Myr gyrochonology-isochrone (gyrochrone) traces the bimodality midpoint up to 6000 K (red solid line), but deviates from the isochrone sharply to $\sim$6200 K (red dotted line). A log-linear extrapolation of the isochrone at 6000K to 6500 K (red dashed line) continues to track the bimodality to hotter temperatures, and roughly follows a line of constant Rossby number. 
}
\label{fig:gyro}
\end{figure*}

%%%%%%%%%%%%%%%%%%%%%%
\section{Extending the Spin-Down Gap}
A bimodal period distribution was first discovered by \citet{mcquillan2013} for \Kepler M dwarfs, who found a dearth of objects with periods around $\sim$25 days. Follow-up work by \citet{mcquillan2014} found this bimodality extended to K dwarfs, up to $T_{eff}\sim5500$. Figure \ref{fig:gyro} shows the rotation period distribution for the final 440 star sample of likely main sequence stars. The bimodality appears to extend smoothly through to the hottest stars in this sample. 

While these periods for field stars were robustly measured by \citet{mcquillan2014} and others, the bimodality did not appear in previous \Kepler work due to the high contamination rate by subgiants for these bluer, hotter stars.  414 rotating stars had acceptable photometric and parallax uncertainties in TGAS, but were culled from this sample for having $M_G$ luminosities higher than the main sequence cut in \S3 above. Note that distributions of $\log g$ values from the Kepler Input Catalog \citep{brown2011a} for both the main sequence and subgiant stars were not statistically different.

The minimum in the bimodal period distribution in Figure \ref{fig:gyro} can be traced using a gyrochonology isochrone (colloquially known as a ``gyrochrone'').  Many different studies have produced competing gyrochrone models, each with unique morphologies at the hot and cool star regimes \citep[e.g.][]{barnes2007,mamajek2008,meibom2011,angus2015}. A 600 Myr gyrochrone from \citet{meibom2011}, converted from $B-V$ color to temperature using the the transformation from \citet{sekiguchi2000}, was determined by eye to approximately trace the period minima from 3500 K to 6000 K. As shown in Figure \ref{fig:gyro}, this model (as with most gyrochronology models) turns down in period sharply for stars hotter than $\sim$6000 K. A log-linear extrapolation of the 600 Myr gyrochrone at 6000 K continues to trace the period bimodality up to 6500 K, and roughly traces a line of constant Rossby number assuming a local convective turnover timescale such as from \citet{barneskim2010}.

%%%%%
\begin{figure}[]
\centering
\includegraphics[width=4.5in]{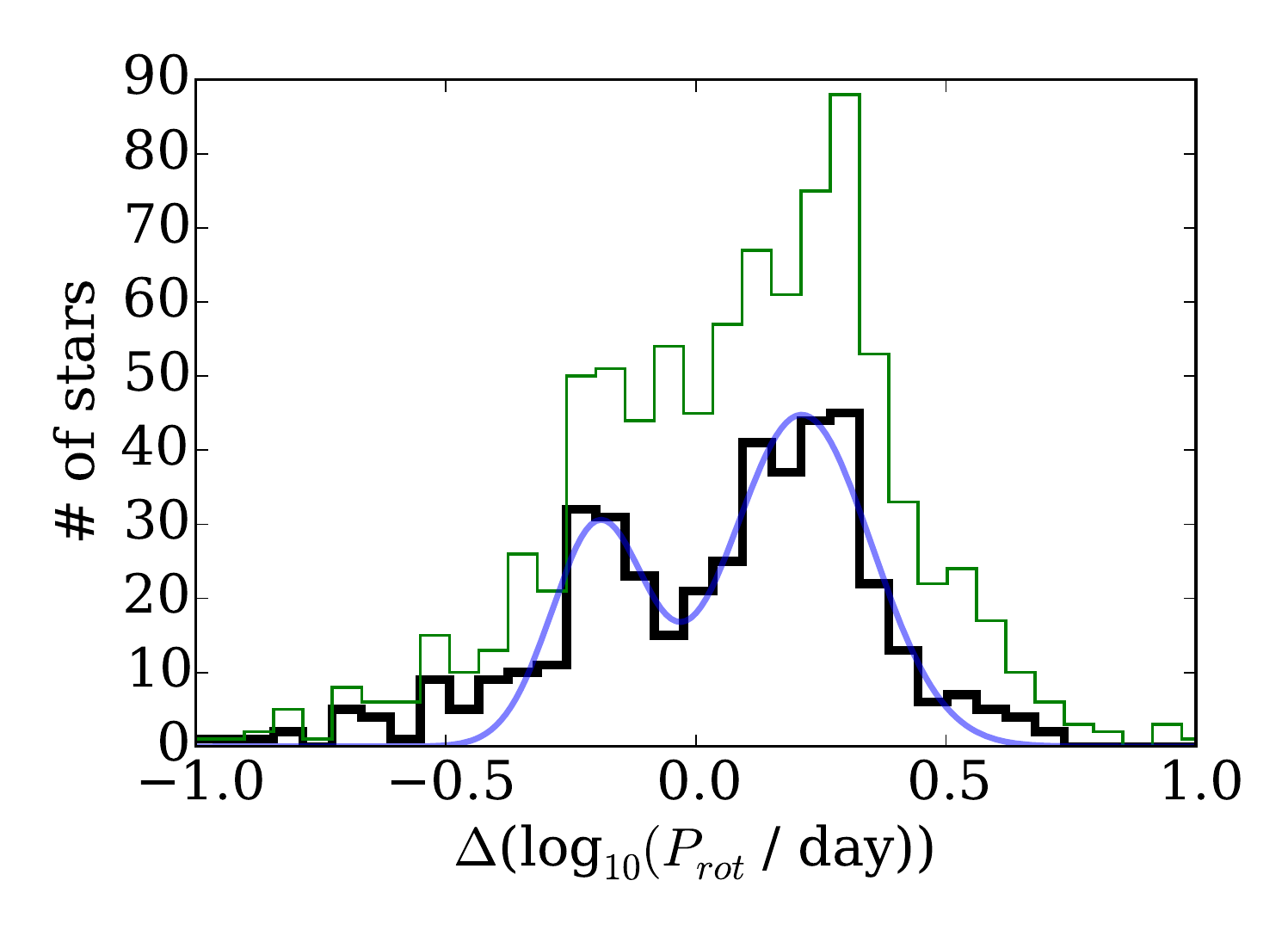}
\caption{Residual of log rotation periods about the \citet{meibom2011} 600 Myr gyrochrone, using the log-linear extrapolation between 6000--6500 K shown in Figure \ref{fig:gyro} \added{for the 440 main sequence selected stars (thick black line) and the initial sample of 894 stars with good TGAS detections (thin green line).}  The bimodal rotation period distribution \added{for main sequence stars} is clear, with peaks at $\Delta \log P_{rot}$ of -0.19 and 0.21 \replaced{days}{dex} from a two-Gaussian model fit (blue line). 
Approximately 50\% more stars are present in the slower, nominally older peak (right).
}
\label{fig:diff}
\end{figure}

The difference (in log period) between the observed rotation and the 600 Myr gyrochrone is shown in Figure \ref{fig:diff}. Despite combining stars of all temperatures, the bimodality is clearly seen in this log period space \added{for the 440 likely main sequence stars}. A two Gaussian model was fit to this data, which found peaks in the two distributions of $-0.19 \pm 0.01$ and $0.21 \pm 0.01$ \replaced{days}{dex}.  262 stars had periods longer than the gyrochrone model (right peak) and 178 slower than the model (left peak). This is in contrast to the overall results from \citet{mcquillan2013} who found nearly equal numbers of M dwarfs in the fast and slow rotating groups, but is in general agreement with their sample of stars with non-zero proper motions. \added{The whole sample of 894 stars with good TGAS detections does not show this bimodality in Figure \ref{fig:diff}, demonstrating the importance of culling subgiants from the sample.}

%%%%%
\begin{figure}[]
\centering
\includegraphics[width=4.5in]{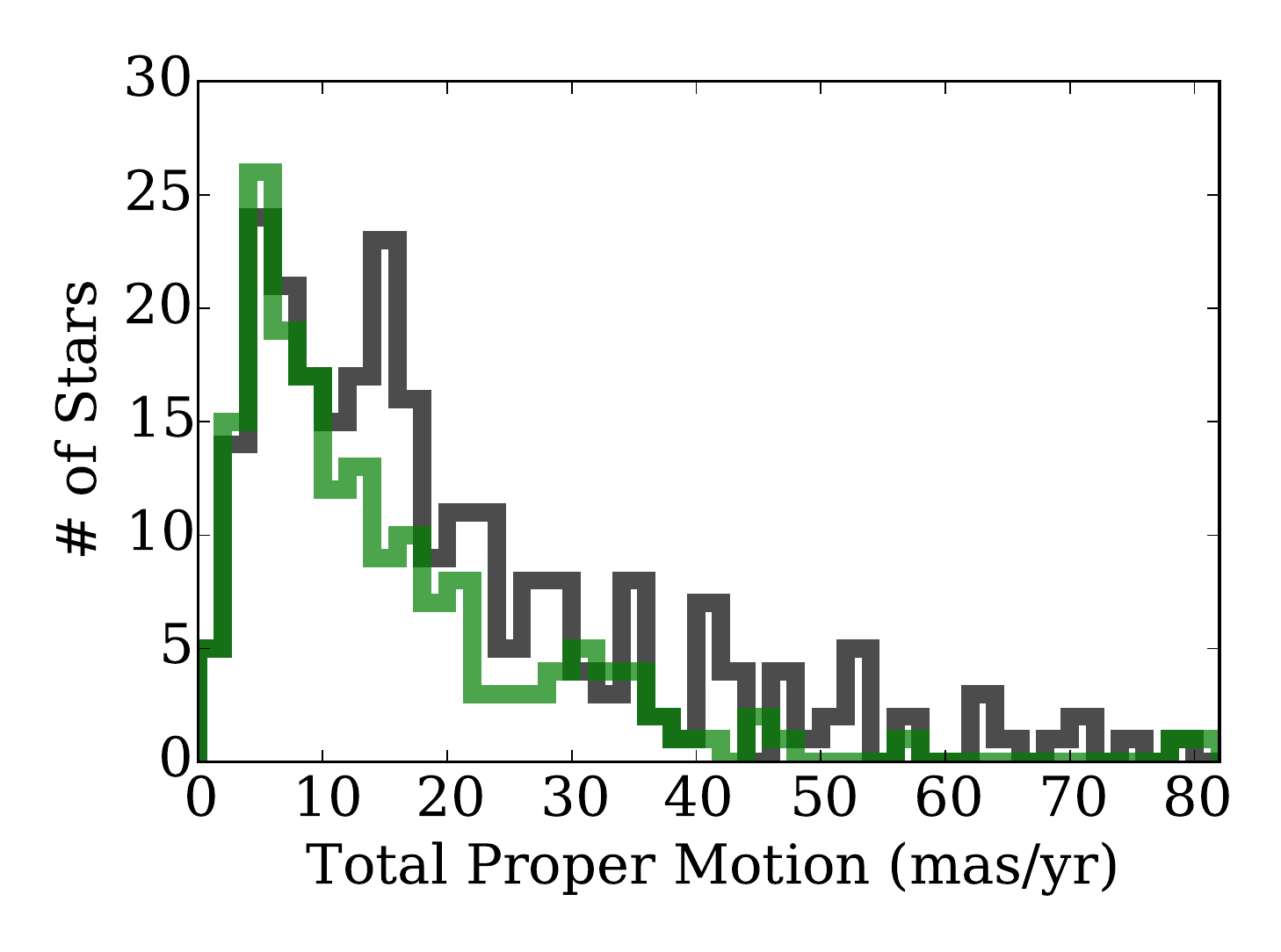}
\includegraphics[width=4.5in]{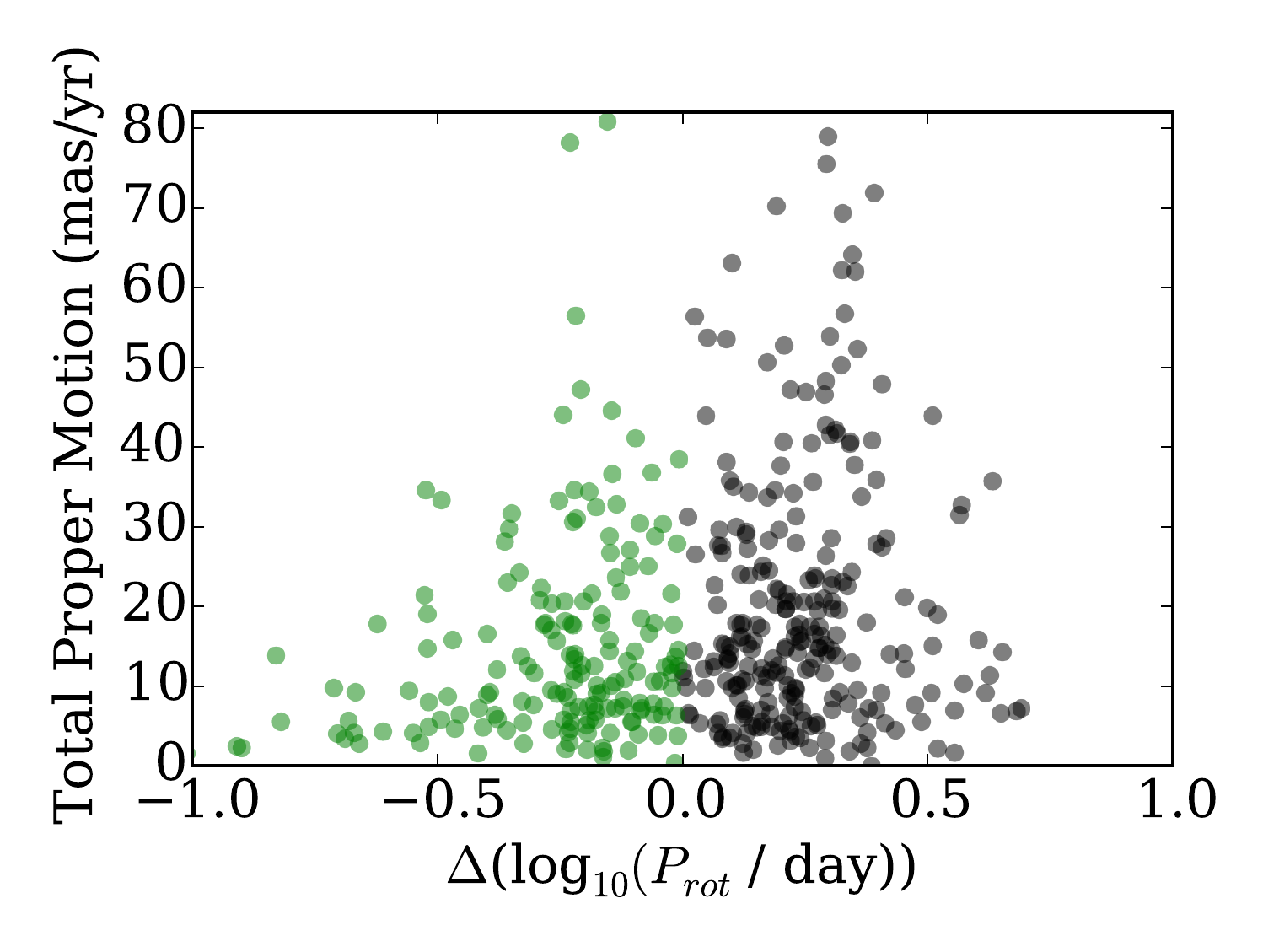}
\caption{Top: Total proper motion distributions for stars above (rotating slower, older stars) the gyrochrone model shown in Figure \ref{fig:gyro} (black), and below (rotating faster, younger stars) the model (green). %\deleted{Median values for the two distributions are shown (vertical dashed lines), which yield a marginal 2.8$\sigma$ difference.}
\added{Bottom: Total proper motion as a function of the residual log rotation period shown in Figure \ref{fig:diff}, with the same color scheme for faster and slower rotating stars as the top panel.}
}
\label{fig:pm}
\end{figure}

The favored explanation for the bimodal period distribution by \citet{mcquillan2013} was an age effect, with nearby stars having a bimodal star formation history. This explanation was bolstered by their observation that stars in the two period groups have differing distributions of proper motions, indicating they belonged to kinematically separate groups. This measurement is replicated in Figure \ref{fig:pm}, which shows the total proper motion distribution for stars above and below the 600 Myr gyrochrone. Stars above the gyrochrone (slower rotators, nominally older) have a median total proper motion of 15.4 mas/yr, while those below (faster rotators, younger)  have a median of 11.3 mas/yr. This difference in kinematics versus rotation period is in the same direction observed by \citet{mcquillan2013}. 
%\deleted{As the typical error in the  Gaia DR1 total proper motion is $\sim$1.4 mas/yr for this sample, this is a statistically marginal (2.8$\sigma$) difference.  Similarly, }
The Kolmogorov-Smirnov statistic for these two samples is 0.14, which does not rule out the null hypothesis that they are drawn from the same distribution. The slower rotating sample (black in Figure \ref{fig:pm}) appears to have a bimodal distribution in proper motion, indicating possibly significant contamination from a younger, lower proper motion population that is consistent with the rapidly rotating stars. \added{Figure \ref{fig:pm} also shows the proper motion as a function of the residual rotation period defined in Figure \ref{fig:diff}. Despite the small sample size, this distribution highlights the complex relationship between ages derived from gyrochronology and galactic kinematics, including bimodal structure for the proper motion of slower rotating stars.} Further investigation of rotation and kinematics of nearby stars using larger samples from future Gaia data releases is needed.

\clearpage
%%%%%%%%%%%%%%%%%%%%%%
\section{Discussion}

Using a combination of data from \Kepler and Gaia DR1, I have explored the rotation period distribution for 440 nearby main sequence stars. A bimodal rotation period distribution has been found in stars with temperatures ranging from 5000 K to 6500 K. This feature matches that found in cooler stars from \Kepler, but was only revealed thanks to the enhanced ability to distinguish dwarfs from subgiants using Gaia data. 
A tenuous difference in the TGAS total proper motion for stars in the fast and slow rotating groups is found, which is in agreement with the findings for cool stars by \citet{mcquillan2013}.

While a definitive explanation for this period bimodality has not been reached, the findings to date seem to favor stellar ages as the cause. In this scenario the star formation history for nearby stars would be dominated by two epochs of star formation, one short event centered at a few hundred Myr, and one long event centered at a few Gyr (slightly younger than the Sun). It is also worth noting that the space volume probed by the TGAS sample investigated here is very similar to that covered by the temperature-selected cool star sample in \citet{mcquillan2013}. The median parallax distance for stars in this work is 285 pc, while the median isochrone distance for the K and M dwarfs is $\sim$216 pc. This points to the period distribution being a localized age artifact. 
Determining how localized this age distribution is, and if it can be confirmed for stars across the HR diagram including giants, is a key goal for future Gaia data releases.

The period bimodality may yet be a manifestation of the ``Vaughan-Preston'' gap observed in chromospheric activity indicators from solar type stars. Such a feature has also been discussed for rotating stars by \citet{kado-fong2016}. Given that the mass range for the bimodality explored here and in \citet{mcquillan2014} covers stars with solar-type dynamos (those having a tachocline, late F through early M) such a model cannot be fully ruled out at this time. Though there have been many rotation studies for cool stars \citep[e.g.][]{irwin2011,newton2016,stelzer2016} too few rotation periods have been measured for stars across the ``fully convective boundary'' ($T_{eff}<3000$ K, spectral type $\sim$M4) to tell if the bimodal period feature continues to cooler temperatures, which would support the age distribution model. 
If the bimodality is due to stars crossing a phase of rapid angular momentum evolution, we would expect to see it in stellar clusters at or near the critical age. The lack of this feature in the clusters observed to date could be due to no cluster being close enough to the critical age, which the gyrochrone in Figure \ref{fig:gyro} shows is near 600 Myr. Further studies of rotation periods for stars in intermediate age open clusters (e.g. the Hyades) may help solve this mystery \citep[e.g.][]{douglas2014}.

Finally, this exploratory work has highlighted the utility of using astrometric data from Gaia combined with detailed light curve statistics from \Kepler to reveal hidden substructure in the properties of field stars. Looking forward to the astrometric precision of future Gaia data releases, this combination will be effective at separating dwarf stars from subgiants for nearly the entire \Kepler and K2 databases, and enable accurate age maps for field stars.
% super useful for determining approximate log g values (separating main sequence from sub-giants), and thus more accurate gyrochronology ages \citep{van-saders2013}. 

%%%%%%%%%%%%%%%%%
\acknowledgments
Special thanks to Kevin Covey, Jennifer van Saders, Sean Matt, Travis Metcalfe, and Aleks Scholz for their helpful discussions that motivated publishing this work.
\added{Thanks to the anonymous referee for their helpful comments that improved this manuscript. Jay Strader provided a helpful reference for the correct notation for transcendental functions, \citet{matta2011}.}
JRAD is supported by an NSF Astronomy and Astrophysics Postdoctoral Fellowship under award AST-1501418. 

This project was developed as part of the 2016 NYC Gaia Sprint, hosted by the Center for Computational Astrophysics at the Simons Foundation in New York City.

This research made use of the cross-match service, and the SIMBAD database, provided by CDS, Strasbourg, France.

This work has made use of data from the European Space Agency (ESA)
mission {\it Gaia} (\url{http://www.cosmos.esa.int/gaia}), processed by
the {\it Gaia} Data Processing and Analysis Consortium (DPAC,
\url{http://www.cosmos.esa.int/web/gaia/dpac/consortium}). Funding
for the DPAC has been provided by national institutions, in particular
the institutions participating in the {\it Gaia} Multilateral Agreement.

%\bibliography{/Users/davenpj3/Dropbox/references.bib}

\begin{thebibliography}{}
\expandafter\ifx\csname natexlab\endcsname\relax\def\natexlab#1{#1}\fi

\bibitem[{{Aigrain} {et~al.}(2015){Aigrain}, {Llama}, {Ceillier}, {Chagas},
  {Davenport}, {Garc{\'{\i}}a}, {Hay}, {Lanza}, {McQuillan}, {Mazeh}, {de
  Medeiros}, {Nielsen}, \& {Reinhold}}]{aigrain2015}
{Aigrain}, S., {Llama}, J., {Ceillier}, T., {et~al.} 2015, \mnras, 450, 3211

\bibitem[{{Angus} {et~al.}(2015){Angus}, {Aigrain}, {Foreman-Mackey}, \&
  {McQuillan}}]{angus2015}
{Angus}, R., {Aigrain}, S., {Foreman-Mackey}, D., \& {McQuillan}, A. 2015,
  \mnras, 450, 1787

\bibitem[{{Barnes}(2007)}]{barnes2007}
{Barnes}, S.~A. 2007, \apj, 669, 1167

\bibitem[{{Barnes} \& {Kim}(2010)}]{barneskim2010}
{Barnes}, S.~A., \& {Kim}, Y.-C. 2010, \apj, 721, 675

\bibitem[{{Bertelli} {et~al.}(1999){Bertelli}, {Bressan}, {Chiosi}, \&
  {Vallenari}}]{bertelli1999}
{Bertelli}, G., {Bressan}, A., {Chiosi}, C., \& {Vallenari}, A. 1999, Baltic
  Astronomy, 8, 271

\bibitem[{{Borucki} {et~al.}(2010){Borucki}, {Koch}, {Basri}, {Batalha},
  {Brown}, {Caldwell}, {Caldwell}, {Christensen-Dalsgaard}, {Cochran},
  {DeVore}, {Dunham}, {Dupree}, {Gautier}, {Geary}, {Gilliland}, {Gould},
  {Howell}, {Jenkins}, {Kondo}, {Latham}, {Marcy}, {Meibom}, {Kjeldsen},
  {Lissauer}, {Monet}, {Morrison}, {Sasselov}, {Tarter}, {Boss}, {Brownlee},
  {Owen}, {Buzasi}, {Charbonneau}, {Doyle}, {Fortney}, {Ford}, {Holman},
  {Seager}, {Steffen}, {Welsh}, {Rowe}, {Anderson}, {Buchhave}, {Ciardi},
  {Walkowicz}, {Sherry}, {Horch}, {Isaacson}, {Everett}, {Fischer}, {Torres},
  {Johnson}, {Endl}, {MacQueen}, {Bryson}, {Dotson}, {Haas}, {Kolodziejczak},
  {Van Cleve}, {Chandrasekaran}, {Twicken}, {Quintana}, {Clarke}, {Allen},
  {Li}, {Wu}, {Tenenbaum}, {Verner}, {Bruhweiler}, {Barnes}, \&
  {Prsa}}]{borucki2010}
{Borucki}, W.~J., {Koch}, D., {Basri}, G., {et~al.} 2010, Science, 327, 977

\bibitem[{{Bressan} {et~al.}(2012){Bressan}, {Marigo}, {Girardi}, {Salasnich},
  {Dal Cero}, {Rubele}, \& {Nanni}}]{bressan2012}
{Bressan}, A., {Marigo}, P., {Girardi}, L., {et~al.} 2012, \mnras, 427, 127

\bibitem[{{Brown} {et~al.}(2011){Brown}, {Latham}, {Everett}, \&
  {Esquerdo}}]{brown2011a}
{Brown}, T.~M., {Latham}, D.~W., {Everett}, M.~E., \& {Esquerdo}, G.~A. 2011,
  \aj, 142, 112

\bibitem[{{Carrasco} {et~al.}(2014){Carrasco}, {Catal{\'a}n}, {Jordi},
  {Tremblay}, {Napiwotzki}, {Luri}, {Robin}, \& {Kowalski}}]{carrasco2014}
{Carrasco}, J.~M., {Catal{\'a}n}, S., {Jordi}, C., {et~al.} 2014, \aap, 565,
  A11

\bibitem[{{Ciardi} {et~al.}(2011){Ciardi}, {von Braun}, {Bryden}, {van Eyken},
  {Howell}, {Kane}, {Plavchan}, {Ram{\'{\i}}rez}, \& {Stauffer}}]{ciardi2011}
{Ciardi}, D.~R., {von Braun}, K., {Bryden}, G., {et~al.} 2011, \aj, 141, 108

\bibitem[{{Creevey} {et~al.}(2013){Creevey}, {Th{\'e}venin}, {Basu}, {Chaplin},
  {Bigot}, {Elsworth}, {Huber}, {Monteiro}, \& {Serenelli}}]{creevey2013}
{Creevey}, O.~L., {Th{\'e}venin}, F., {Basu}, S., {et~al.} 2013, \mnras, 431,
  2419

\bibitem[{{do Nascimento} {et~al.}(2012){do Nascimento}, {da Costa}, \&
  {Castro}}]{donascimento2012}
{do Nascimento}, J.-D., {da Costa}, J.~S., \& {Castro}, M. 2012, \aap, 548, L1

\bibitem[{{Douglas} {et~al.}(2014){Douglas}, {Ag{\"u}eros}, {Covey}, {Bowsher},
  {Bochanski}, {Cargile}, {Kraus}, {Law}, {Lemonias}, {Arce}, {Fierroz}, \&
  {Kundert}}]{douglas2014}
{Douglas}, S.~T., {Ag{\"u}eros}, M.~A., {Covey}, K.~R., {et~al.} 2014, \apj,
  795, 161

\bibitem[{{Gaensicke} {et~al.}(2015){Gaensicke}, {Tremblay}, {Barstow}, {Bono},
  {Burleigh}, {Casewell}, {Dhillon}, {Farihi}, {Garcia-Berro}, {Geier},
  {Gentile-Fusillo}, {Hermes}, {Hollands}, {Istrate}, {Jordan}, {Knigge},
  {Manser}, {Marsh}, {Nelemans}, {Pala}, {Raddi}, {Tauris}, {Toloza}, {Veras},
  {Werner}, \& {Wilson}}]{gaensicke2015}
{Gaensicke}, B., {Tremblay}, P.-E., {Barstow}, M., {et~al.} 2015, ArXiv
  e-prints, arXiv:1506.02653

\bibitem[{{Gaia Collaboration}(2016)}]{gaia}
{Gaia Collaboration}. 2016, ArXiv e-prints, arXiv:1609.04153

\bibitem[{{Hernandez} {et~al.}(2000){Hernandez}, {Valls-Gabaud}, \&
  {Gilmore}}]{hernandez2000}
{Hernandez}, X., {Valls-Gabaud}, D., \& {Gilmore}, G. 2000, \mnras, 316, 605

\bibitem[{{Irwin} {et~al.}(2011){Irwin}, {Berta}, {Burke}, {Charbonneau},
  {Nutzman}, {West}, \& {Falco}}]{irwin2011}
{Irwin}, J., {Berta}, Z.~K., {Burke}, C.~J., {et~al.} 2011, \apj, 727, 56

\bibitem[{{Kado-Fong} {et~al.}(2016){Kado-Fong}, {Williams}, {Mann}, {Berger},
  {Burgett}, {Chambers}, {Huber}, {Kaiser}, {Kudritzki}, {Magnier},
  {Wainscoat}, \& {Waters}}]{kado-fong2016}
{Kado-Fong}, E., {Williams}, P.~K.~G., {Mann}, A.~W., {et~al.} 2016, ArXiv
  e-prints, arXiv:1608.00978

\bibitem[{{Lindegren} {et~al.}(2016){Lindegren}, {Lammers}, {Bastian},
  {Hern{\'a}ndez}, {Klioner}, {Hobbs}, {Bombrun}, {Michalik}, {Ramos-Lerate},
  {Butkevich}, {Comoretto}, {Joliet}, {Holl}, {Hutton}, {Parsons},
  {Steidelm{\"u}ller}, {Abbas}, {Altmann}, {Andrei}, {Anton}, {Bach},
  {Barache}, {Becciani}, {Berthier}, {Bianchi}, {Biermann}, {Bouquillon},
  {Bourda}, {Br{\"u}semeister}, {Bucciarelli}, {Busonero}, {Carlucci},
  {Casta{\~n}eda}, {Charlot}, {Clotet}, {Crosta}, {Davidson}, {de Felice},
  {Drimmel}, {Fabricius}, {Fienga}, {Figueras}, {Fraile}, {Gai}, {Garralda},
  {Geyer}, {Gonz{\'a}lez-Vidal}, {Guerra}, {Hambly}, {Hauser}, {Jordan},
  {Lattanzi}, {Lenhardt}, {Liao}, {L{\"o}ffler}, {McMillan}, {Mignard}, {Mora},
  {Morbidelli}, {Portell}, {Riva}, {Sarasso}, {Serraller}, {Siddiqui}, {Smart},
  {Spagna}, {Stampa}, {Steele}, {Taris}, {Torra}, {van Reeven}, {Vecchiato},
  {Zschocke}, {de Bruijne}, {Gracia}, {Raison}, {Lister}, {Marchant},
  {Messineo}, {Soffel}, {Osorio}, {de Torres}, \& {O'Mullane}}]{gaia_dr1}
{Lindegren}, L., {Lammers}, U., {Bastian}, U., {et~al.} 2016, ArXiv e-prints,
  arXiv:1609.04303

\bibitem[{{Lurie} {et~al.}(2015){Lurie}, {Davenport}, {Hawley}, {Wilkinson},
  {Wisniewski}, {Kowalski}, \& {Hebb}}]{lurie2015}
{Lurie}, J.~C., {Davenport}, J.~R.~A., {Hawley}, S.~L., {et~al.} 2015, \apj,
  800, 95

\bibitem[{{Mamajek} \& {Hillenbrand}(2008)}]{mamajek2008}
{Mamajek}, E.~E., \& {Hillenbrand}, L.~A. 2008, \apj, 687, 1264

\bibitem[{{Mann} {et~al.}(2012){Mann}, {Gaidos}, {L{\'e}pine}, \&
  {Hilton}}]{mann2012}
{Mann}, A.~W., {Gaidos}, E., {L{\'e}pine}, S., \& {Hilton}, E.~J. 2012, \apj,
  753, 90

\bibitem[{{Massarotti}(2008)}]{massarotti2008}
{Massarotti}, A. 2008, \aj, 135, 2287

\bibitem[{{Matta} {et~al.}(2011){Matta}, {Massa}, {Gubskaya}, \&
  {Knoll}}]{matta2011}
{Matta}, C.~F., {Massa}, L., {Gubskaya}, A.~V., \& {Knoll}, E. 2011, Journal of
  Chemical Education, 88, 67

\bibitem[{{McQuillan} {et~al.}(2013){McQuillan}, {Aigrain}, \&
  {Mazeh}}]{mcquillan2013}
{McQuillan}, A., {Aigrain}, S., \& {Mazeh}, T. 2013, \mnras, 432, 1203

\bibitem[{{McQuillan} {et~al.}(2014){McQuillan}, {Mazeh}, \&
  {Aigrain}}]{mcquillan2014}
{McQuillan}, A., {Mazeh}, T., \& {Aigrain}, S. 2014, \apjs, 211, 24

\bibitem[{{Meibom} {et~al.}(2011){Meibom}, {Barnes}, {Latham}, {Batalha},
  {Borucki}, {Koch}, {Basri}, {Walkowicz}, {Janes}, {Jenkins}, {Van Cleve},
  {Haas}, {Bryson}, {Dupree}, {Furesz}, {Szentgyorgyi}, {Buchhave}, {Clarke},
  {Twicken}, \& {Quintana}}]{meibom2011}
{Meibom}, S., {Barnes}, S.~A., {Latham}, D.~W., {et~al.} 2011, \apjl, 733, L9

\bibitem[{{Newton} {et~al.}(2016){Newton}, {Irwin}, {Charbonneau},
  {Berta-Thompson}, {Dittmann}, \& {West}}]{newton2016}
{Newton}, E.~R., {Irwin}, J., {Charbonneau}, D., {et~al.} 2016, \apj, 821, 93

\bibitem[{{Reinhold} {et~al.}(2013){Reinhold}, {Reiners}, \&
  {Basri}}]{reinhold2013}
{Reinhold}, T., {Reiners}, A., \& {Basri}, G. 2013, \aap, 560, A4

\bibitem[{{Sekiguchi} \& {Fukugita}(2000)}]{sekiguchi2000}
{Sekiguchi}, M., \& {Fukugita}, M. 2000, \aj, 120, 1072

\bibitem[{{Skrutskie} {et~al.}(2006){Skrutskie}, {Cutri}, {Stiening},
  {Weinberg}, {Schneider}, {Carpenter}, {Beichman}, {Capps}, {Chester},
  {Elias}, {Huchra}, {Liebert}, {Lonsdale}, {Monet}, {Price}, {Seitzer},
  {Jarrett}, {Kirkpatrick}, {Gizis}, {Howard}, {Evans}, {Fowler}, {Fullmer},
  {Hurt}, {Light}, {Kopan}, {Marsh}, {McCallon}, {Tam}, {Van Dyk}, \&
  {Wheelock}}]{2mass}
{Skrutskie}, M.~F., {Cutri}, R.~M., {Stiening}, R., {et~al.} 2006, \aj, 131,
  1163

\bibitem[{{Skumanich}(1972)}]{skumanich1972}
{Skumanich}, A. 1972, \apj, 171, 565

\bibitem[{{Stassun} \& {Torres}(2016)}]{stassun2016}
{Stassun}, K.~G., \& {Torres}, G. 2016, ArXiv e-prints, arXiv:1609.05390

\bibitem[{{Stelzer} {et~al.}(2016){Stelzer}, {Damasso}, {Scholz}, \&
  {Matt}}]{stelzer2016}
{Stelzer}, B., {Damasso}, M., {Scholz}, A., \& {Matt}, S.~P. 2016, \mnras, 463,
  1844

\bibitem[{{Tayar} {et~al.}(2015){Tayar}, {Ceillier},
  {Garc{\'{\i}}a-Hern{\'a}ndez}, {Troup}, {Mathur}, {Garc{\'{\i}}a}, {Zamora},
  {Johnson}, {Pinsonneault}, {M{\'e}sz{\'a}ros}, {Allende Prieto}, {Chaplin},
  {Elsworth}, {Hekker}, {Nidever}, {Salabert}, {Schneider}, {Serenelli},
  {Shetrone}, \& {Stello}}]{tayar2015}
{Tayar}, J., {Ceillier}, T., {Garc{\'{\i}}a-Hern{\'a}ndez}, D.~A., {et~al.}
  2015, \apj, 807, 82

\bibitem[{{van Saders} {et~al.}(2016){van Saders}, {Ceillier}, {Metcalfe},
  {Silva Aguirre}, {Pinsonneault}, {Garc{\'{\i}}a}, {Mathur}, \&
  {Davies}}]{van-saders2016}
{van Saders}, J.~L., {Ceillier}, T., {Metcalfe}, T.~S., {et~al.} 2016, \nat,
  529, 181

\bibitem[{{Vaughan} \& {Preston}(1980)}]{vaughan1980}
{Vaughan}, A.~H., \& {Preston}, G.~W. 1980, \pasp, 92, 385

\end{thebibliography}

\end{document}